\documentclass[sigconf]{acmart}

\usepackage{graphicx}
\usepackage{algorithm}
\usepackage{algorithmic}
\usepackage{booktabs}
\usepackage{array}
\usepackage{multirow}
\usepackage{morefloats} % allow many full-width figure*/table* floats

\setcopyright{acmlicensed}        % TODO: exact value from eRights email
\acmDOI{10.1145/XXXXXXX.XXXXXXX}  % TODO: from eRights email
\acmISBN{978-1-4503-XXXX-X/26/09} % TODO: from eRights email
\acmConference[OARS @ RecSys '26]{RecSys 2026 Workshop on Online and
  Adaptive Recommender Systems}{September 28, 2026}{Minneapolis, MN, USA}
\acmYear{2026}


\begin{document}

\title{Real-Time Hard Negative Sampling via LLM-based Clustering for Large-Scale Two-Tower Retrieval}

% Single-blind review: the author block is shown in both the review and
% camera-ready PDFs.
\author{Ivan Ji}
\authornote{Three authors contributed equally to this research.}
\affiliation{\institution{Meta}\city{Menlo Park}\state{CA}\country{USA}}
\email{ivanji@meta.com}

\author{Liuyi Hu}
\authornotemark[1]
\affiliation{\institution{Meta}\city{Menlo Park}\state{CA}\country{USA}}
\email{liuyihu@meta.com}

\author{Harrison (Zihao) Zhao}
\authornotemark[1]
\affiliation{\institution{Meta}\city{Menlo Park}\state{CA}\country{USA}}
\email{harrisonzhao@meta.com}

\author{Lei Huang}
\affiliation{\institution{Meta}\city{Menlo Park}\state{CA}\country{USA}}
\email{leihuang@meta.com}

\author{Qunshu Zhang}
\affiliation{\institution{Meta}\city{Menlo Park}\state{CA}\country{USA}}
\email{qunshuzhang@meta.com}

\author{Max (Xiangjun) Fan}
\affiliation{\institution{Meta}\city{Menlo Park}\state{CA}\country{USA}}
\email{maxfan@meta.com}

\author{Aameek Singh}
\affiliation{\institution{Meta}\city{Menlo Park}\state{CA}\country{USA}}
\email{aameeksingh@meta.com}

\renewcommand{\shortauthors}{Ji et al.}

\begin{abstract}
% REVISED: tightened from ~245 to ~190 words. Removed the duplicated method
% definition and two of the three "industrial scale" repetitions. Added the
% qualifier "in our production deployment" to scope the LLM claim accurately
% (R2/R3 flagged the LLM framing as the least-validated part), and surfaced the
% 14-day A/B test. To drop the qualifier, delete only those four words.
The two-tower model is widely used in the retrieval stage of large-scale recommendation systems, where training typically relies on in-batch and/or out-of-batch negative sampling. These methods, however, tend to produce easy negatives that the model learns quickly and that provide little training signal. This paper proposes a self-supervised, cluster-based hard negative sampling technique that draws negatives from the same semantic cluster as the positive item; in our production deployment the clusters are derived from large language model (LLM) based multimodal content representations, so that intra-cluster items are genuinely similar and yield informative negatives. To make this deployable at industrial scale, we realize the technique in a real-time, end-to-end framework that maintains a live in-memory item pool and draws cluster-based negatives from it on the fly via global out-of-batch sampling (GOOBS). The framework integrates directly into production two-tower training and serving and scales to billions of training examples with minimal computational overhead. Experiments on four public datasets and a 14-day online A/B test in a large-scale production system show that the proposed technique outperforms widely used industry methods, while also helping to break recommendation feedback loops and substantially reducing popularity bias.
\end{abstract}


\maketitle

\section{Introduction}
\label{section:intro}

In the era of big data, large-scale recommendation systems have become increasingly important in various applications, including e-commerce, social media, and entertainment. These systems aim to provide users with personalized recommendations that cater to their interests and preferences. However, designing an efficient and effective recommendation system is a challenging task due to the vast number of candidates eligible for selection.
To address this challenge, modern recommendation systems typically adopt a multi-stage design, consisting of retrieval and ranking stages \cite{covington2016deep, liu2017cascade, chen2017efficient}. The retrieval stage aims to retrieve a small subset of relevant candidates from a large corpus, while the ranking stages further refine the selection to provide the most relevant recommendations. The two-tower model design is widely used as retrieval models in large scale applications due to its serving efficiency \cite{covington2016deep, huang2020embedding}.

Retrieval model training is often formulated as an extreme classification problem in which negative samples play a critical role. Various sampling-based techniques have been proposed to enhance training efficiency \cite{bengio2008adaptive, bengio2003quick, covington2016deep}. The widely used sampling techniques are in-batch negative sampling \cite{hidasi2015session, gillick2019learning} and out-of-batch (OOB) negative sampling and also mixed negative sampling \cite{yang2020mixed, hidasi2018recurrent}. However, in-batch negatives are constrained by the mini-batch size, which can cause recommendation bias and limited exposure to a diverse corpus during training. Meanwhile, random OOB negatives are often too easy for the model to learn, especially when the OOB item pool is large and diverse.

Besides, retrieval models typically use user engagement such as click as positives. However, whether user clicks or not depends on what item we surface to the user which is controlled by the multi-stage system. The strong feedback loop can result in popularity bias in the recommendation \cite{chen2020bias, canamares2018should, morik2020controlling, oosterhuis2021computationally}.

To solve the problems of negatives being too easy and also popularity bias, this paper proposes a self-supervised cluster-based hard negative sampling technique for the two-tower model, together with a practical real-time framework that makes it deployable at industrial scale. The technique generates hard negatives from item clusters on the fly during training, while the framework integrates directly into a production two-tower training and serving pipeline, scaling to billions of examples with minimal overhead. On public datasets, experiments demonstrate that the proposed negative sampling technique yields significant performance improvements, particularly on large-scale datasets such as Amazon Reviews, which feature a vast number of unique items. When deployed as a retrieval source in a production recommendation system comprising multiple candidate sources, the proposed model achieves a $+53\%$ CTR improvement on its served impressions relative to the prior source model. The technique also helps mitigate popularity bias.

The contributions of this paper are summarized below.
\begin{itemize}
    \item \textbf{Technique.} A self-supervised cluster-based hard negative sampling method for two-tower models that draws negatives from the same semantic cluster, using clusters derived from LLM-based multimodal content representations.
    \item \textbf{A practical real-time realization} that makes cluster-based sampling deployable in production two-tower training and serving, generating negatives on the fly with negligible overhead.
    \item Experiments on public and industrial datasets demonstrate consistent gains over the widely used in-batch and out-of-batch negative sampling methods.
    \item A production deployment demonstrates that the approach also mitigates popularity bias at industrial scale.
\end{itemize}

\section{Related Work}
\subsection{Multi-stage System}
For large scale recommendation systems, there are millions of candidates eligible for selection. Given the latency constraint, it is infeasible to rank all the candidates with a complicated ranking model. To balance recommendation efficiency and effectiveness, modern recommendation system typically adopts a multi-stage design: retrieval and ranking. The retrieval stage aims to retrieve a small subset from a large corpus. Then several ranking stages are deployed to rerank the retrieved subset. Such multi-stage system has been widely used in both industry and academia \cite{covington2016deep, liu2017cascade, chen2017efficient}.

\subsection{Two-Tower Model}
The Two-Tower model, a variant of the neural network architecture, has been a powerful tool in recommendation systems since \cite{huang2013learning} introduced it. It is widely used in the retrieval stage of industry applications \cite{covington2016deep, huang2020embedding}. One of the towers is typically used to model user interactions, and the other is used to model item characteristics. The user and item are represented by the embedding from the user and item tower. Then the retrieval problem is converted into a nearest neighbor (NN) search problem in the embedding space. The biggest advantage for this architecture is execution efficiency since the item embedding for the entire corpus can be precomputed and indexed so that online inference in real time is not required.

\subsection{Negative Sampling}
Retrieval model training is often formulated as an extreme classification problem, with various sampling-based techniques being proposed to enhance training efficiency \cite{bengio2008adaptive, bengio2003quick, covington2016deep}.

\subsubsection{In-batch Negative Sampling}
It treats positive items of other users in the same mini-batch as negative items, rather than selecting from the corpus. It is commonly used owing to its time and memory efficiency \cite{hidasi2015session, gillick2019learning}. However, in-batch negatives are inherently limited by the mini-batch size. This can cause recommendation bias and limits the model's exposure to diverse candidates during training. To mitigate this, there are several variants being proposed. For example, \cite{wang2021cross} has proposed cross-batch negative sampling which takes advantage of the encoded items' embeddings from recent mini-batches to boost model training.

\subsubsection{LogQ Correction}
LogQ correction is a technique used in large-scale recommendation systems to correct for bias introduced by sampled softmax during training, where popular items are more likely to be sampled as negative examples. It has been well adopted in the industry by companies like Google \cite{yi2019sampling, yang2020mixed}, ByteDance \cite{yan2024trinity}, Kuaishou \cite{liu2024kuaiformer}. It works by adjusting the model's logits (raw output scores) by subtracting the log-probability of a negative item's occurrence in the training batch, thereby penalizing popular items less and improving the model's ability to recall less frequent items. This correction helps ensure that the model learns from true preference signals rather than statistical artifacts of the sampling process.

\subsubsection{Out-of-batch (OOB) Negative Sampling}
Out-of-batch negative samples are selected from a pool of items that are not present in the current mini-batch of data being processed.
By sampling from the items out of the current mini-batch, it provides a wider range of negative samples, potentially leading to better model generalization as the model encounters a broader variety of items that could be considered ``not relevant''. It also makes the training and serving more consistent since during serving time, the model needs to retrieve relevant items from the whole corpus instead of just items from the mini-batch. However, it can be computationally more expensive compared to in-batch sampling as it might require accessing and processing data outside the current mini-batch. Also, random OOB negative samples are very easy negatives for the model to distinguish. To overcome this obstacle, many different hard negative sampling methods have been proposed in research, such as Dynamic Negative Sampling \cite{zhang2013optimizing}, Adaptive Sampling \cite{chen2022learning, wang2017irgan}, and robust negative sampling \cite{ding2020simplify}. But they are not widely adopted for large scale industrial applications due to the computational cost. In contrast, the method proposed in this paper samples hard negatives from semantic clusters in real time, requiring neither a global approximate-nearest-neighbor index (as in ANCE~\cite{xiong2020approximate}) nor reliance on batch recency (as in cross-batch sampling~\cite{wang2021cross}), which makes it well suited to industrial-scale streaming training.

\subsubsection{Mixed Negative Sampling}
Mixed Negative Sampling uses a mixture of in-batch and out-of-batch sampled negatives to tackle the selection bias of implicit user feedback. This technique is also widely used in various applications \cite{yang2020mixed, hidasi2018recurrent}.

\subsection{Popularity Bias}
The ideal state of recommendation system is to help find the most relevant items to users in a personalized way. However, in real applications, we often see popularity bias in the recommendation.
Popularity bias means that the recommendation system tends to recommend rather popular items to users at the expense of less popular items that users may find relevant.
The problem of popularity bias in recommendation systems has garnered significant attention from researchers over the past decade due to its practical importance \cite{chen2020bias, canamares2018should, morik2020controlling, oosterhuis2021computationally}.

Overall, popularity bias in the recommendation system has the following negative effects: (1) making users interact with a limited number of items and therefore hurting user experiences; (2) no exploration on the long tail items to learn the embedding; (3) causing strong system feedback loop that is hard to break since already popular items will receive more exposures and become even more popular.

To mitigate, some methods directly try to correct the bias in the model training itself \cite{abdollahpouri2017controlling, steck2011item, wei2021model}, while others try to correct in a post-processing strategy via adjusting the predictions of the model with a bias factor \cite{abdollahpouri2019managing, zhu2021popularity}. Another typical approach is to balance the popular and unpopular items in the exposure data by assigning weights that are inversely proportional to item popularity in the loss function \cite{steck2011item}.

\section{Methodology}
\subsection{Problem Formulation}

We denote the labeled samples as $\{x_i, y_i, r_i\}_{i=1}^{n}$ where $x_i$ indicates the user side information, $y_i$ indicates the item side information and $r_i$ indicates the label for $i$th example pair $(x_i, y_i)$.

The retrieval model is typically modeled as an extreme multi-class classifier. The goal of the model is to predict the probability of user $x_i$ engaging with item $y_i$
\begin{displaymath}
  P(y_i | x_i; \theta) = \frac{e^{s(x_i, y_i | \theta)}}{\sum_{j \in \mathcal{I}} e^{s(x_i, y_j | \theta)}}
\end{displaymath}
where $\mathcal{I}$ is the eligible item set, $s(x_i, y_i | \theta)$ is some function based on $x_i$ and $y_i$ and $\theta$ is the model parameter. For example, in the basic two-tower model set-up, $s(x_i, y_i) = v_i^T u_i$ where $v_i$ and $u_i$ are the embedding output from the user and item tower respectively. In the industrial applications, the cardinality of the item set $\mathcal{I}$ is at least on the order of millions.

Considering the widely adopted cross-entropy loss, the loss function to optimize is

\begin{displaymath}
\mathcal{L} (\{x_i, y_i, r_i\}_{i=1}^{n}) = - \frac{1}{n} \sum_{i=1}^n  r_i \cdot \log (P(y_i | x_i; \theta))
\end{displaymath}

\subsection{Self-Supervised Cluster-based Negative Sampling}
To enhance the training efficiency, the proposed method relies on a self-supervised cluster-based negative sampling technique to sample hard negatives from the item pool $\mathcal{I}$.
First, clustering is performed on the item pool $\mathcal{I}$ and each $y_j$ is assigned a cluster id. The clustering can be based on various dimensions, e.g.\ item category, item topics etc. Then for each example $(x_i, y_i)$ in the labeled samples, $y_{i1}^{-n}, y_{i2}^{-n}, \ldots, y_{iK}^{-n}$ negatives are sampled from the same cluster as $y_i$. $\{y_{ik}^{-n}\}$ are hard negatives for the $(x_i, y_i)$ because they are similar to $y_i$ in some way.

For each example the cross-entropy loss is minimized for the true label and the sampled negative classes, i.e.

\begin{displaymath}
\resizebox{\columnwidth}{!}{$\displaystyle
\mathcal{L} (\{x_i, y_i, r_i\}, \{x_i, y_{ik}^{-n}, 0\}_{k=1}^K) = -  r_i \cdot \log \left(\frac{e^{s(x_i, y_i | \theta)}}{\sum_{j \in \{y_i\} \cup \{y_{ik}^{-n}\}_{k=1}^{K}} e^{s(x_i, y_j | \theta)}}\right)
$}
\end{displaymath}

\subsection{A Gradient Perspective on Cluster-based Negatives}\label{sec:theory}
The efficacy of cluster-based negative sampling can be understood
intuitively through the lens of contrastive learning and gradient
analysis. In a standard two-tower model trained
with InfoNCE or softmax cross-entropy loss, the gradient
% REVISED: notation aligned with Section 3.1 ($x_i$ user, $y_i$ positive item,
% $y_{ik}^{-n}$ sampled negative). The previous $x_u$ / $x_j^-$ were undefined.
contribution of a sampled negative $y_{ik}^{-n}$ with respect to the
user anchor $x_i$ is proportional to its exponentiated similarity
$\exp\!\big(s(x_i, y_{ik}^{-n} \mid \theta)\big)$. Uniformly sampled negatives typically
reside far from the anchor in the embedding space, resulting in
vanishingly small similarities and, consequently, near-zero
gradients that provide minimal learning signal.

By sampling negatives from the same semantic cluster as the
positive item $y_i$, the proposed method guarantees that $y_{ik}^{-n}$ shares latent
characteristics with $y_i$, ensuring a higher
similarity score $s(x_i, y_{ik}^{-n} \mid \theta)$ prior to full convergence. This
pushes the negative sample closer to the model's current decision
boundary. Consequently, the gradient magnitude is significantly
larger, forcing the model to learn fine-grained distinctions
between intra-cluster items rather than relying on trivial
inter-cluster differences. This approach aligns with the
theoretical findings in Approximate Nearest Neighbor Negative
Contrastive Estimation (ANCE)~\cite{xiong2020approximate}, which
demonstrates that negatives drawn from the local neighborhood of
the query yield higher gradient norms and better approximate the
oracle importance sampling distribution. This analysis also predicts a
failure mode we observe empirically: because in-cluster negatives carry
disproportionately large gradients, sampling too many of them per positive
over-weights the contrastive term and destabilizes training. Indeed, varying
the number of in-cluster hard negatives per positive from 1 to 64, we find
that a small number performs best while larger counts degrade performance.

\subsection{LLM-based Cluster Generation} \label{llm}

In the cluster-based negative sampling process, selecting the appropriate clusters is crucial.
Traditionally, media genres or categories have been used as clustering options, as discussed in
Section~\ref{public_data}. However, for the industrial application described in this paper,
item clusters are derived from an in-house multimodal content-embedding model that captures
semantic relationships across text, image, audio, and video, going beyond surface-level genre or category.

\begin{figure*}[t]
  \centering
  \includegraphics[width=0.62\textwidth]{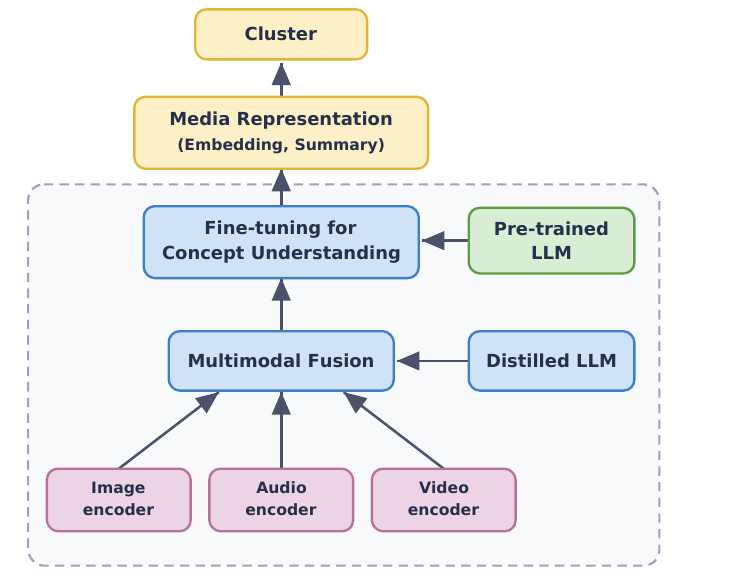}
  \caption{LLM-based cluster generation}
  \label{fig:interestFM}
\end{figure*}

The content embedding model is constructed as a set of fine-tuned encoders built on top of a
large language model (LLM), designed to produce semantically rich item representations across
text, image, audio, and video modalities. Figure~\ref{fig:interestFM} illustrates the workflow, which
includes the following steps:

\begin{itemize}
    \item \textit{Pre-training}: The pre-trained LLM is utilized to achieve robust general language understanding. This foundational step ensures that the model captures complex semantic relationships that are often missed by traditional clustering methods based solely on surface-level features like genre or category.
    \item \textit{Multimodal Encoder}: This module processes diverse input types, including text, images, audio, and videos, using a transformer-based architecture to encode them into fixed-size vector representations. This multimodal capability allows for a more nuanced and comprehensive understanding of media content, surpassing the limitations of traditional clustering that typically relies on single-modal data.
    \item \textit{Fine-tuning}: The model is then fine-tuned to adapt to specific tasks, ensuring that the clusters generated are highly relevant and contextually informed. This fine-tuning process allows the model to capture subtle distinctions and patterns in user interests that traditional clustering methods might overlook.
\end{itemize}
Given these embeddings, item clusters are formed by applying $k$-means to the content-embedding space; each item is assigned to its nearest centroid, and newly encoded items are assigned online as they arrive. Because items are grouped by semantic proximity in this LLM-derived representation---rather than by surface genre or category---intra-cluster items tend to be genuinely similar across modalities, which yields harder and more informative negatives for cluster-based sampling.

The clustering granularity, controlled by the number of clusters $k$, plays a crucial role. Overly fine-grained clusters increase the risk of false negatives---items so similar to the positive that they are effectively relevant---whereas overly coarse clusters dilute the hardness of the sampled negatives. We therefore adopt a relatively coarse granularity in the production deployment.

The risk of sampling false negatives from within a cluster is small: a user is genuinely interested in at most thousands of items out of billions, so within a single cluster of order $10^4$ items the fraction truly relevant to a given user---and thus false negatives if sampled---is well below $1\%$, and uniform random sampling further disperses these rare collisions across training steps, making their effect on the gradient negligible.

\section{A Scalable Real-Time Sampling System}

\begin{figure*}[t]
  \centering
  \includegraphics[width=0.7\textwidth]{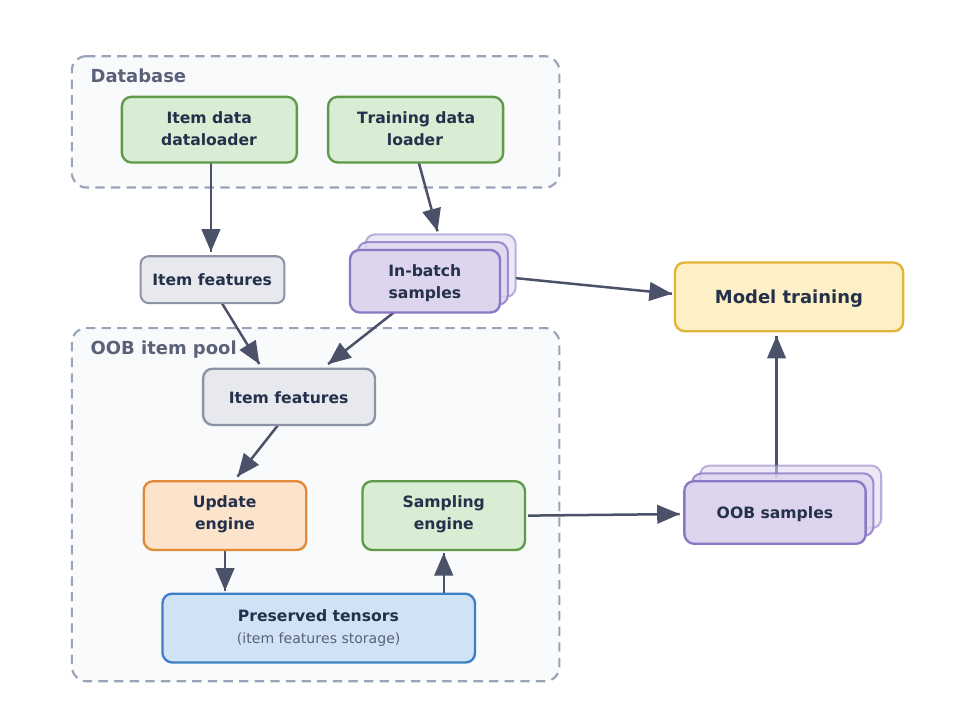}
  \caption{The real-time cluster-based negative sampling system}
  \label{fig:sampling_fw}
\end{figure*}

To deploy this technique at scale, we realize it in a real-time, end-to-end framework that performs cluster-based negative sampling in production, integrating into two-tower training and serving while handling billions of training examples with minimal computational overhead. The framework performs global out-of-batch sampling (GOOBS): as illustrated in Figure \ref{fig:sampling_fw}, it maintains an item pool that stores tensors for OOB samples. The maximum number of items is predefined, and each item is assigned to a ``slot,'' which is a set of tensors storing the item's features.

During training, in-batch samples are passed through an update engine that employs a customized hashing function to determine the slot where the item should be stored. In parallel, a sampling engine takes in-batch items' cluster IDs as input to sample corresponding OOB samples from the item pool. These OOB samples are then combined with in-batch samples for training.

To ensure a high sampling hit rate during the early stages of training, the OOB item pool is pre-loaded with item features from an item data table derived from previous training data. Although this data may be slightly delayed compared to the latest training data, in-batch training samples capture the latest available items and continuously update the OOB item pool to maintain its freshness.

\begin{figure*}[t]
  \centering
  \includegraphics[width=0.6\textwidth]{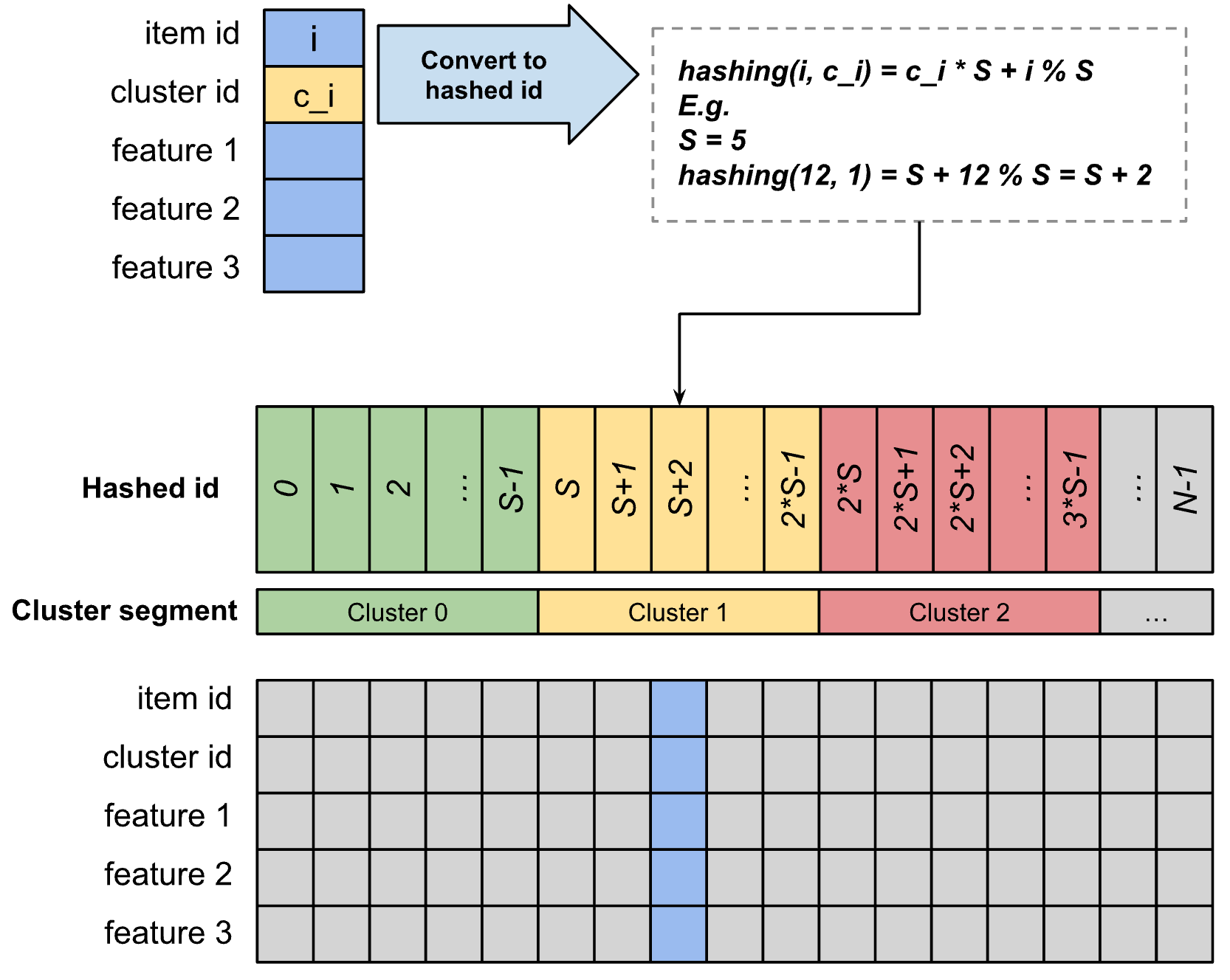}
  \caption{Cluster-based pool update}
  \label{fig:pool_update}
\end{figure*}

The core functionality of the cluster-based negative sampling framework is embedded in the update engine and the sampling engine. As shown in Figure \ref{fig:pool_update} and Algorithm \ref{alg:update}, during the update process, the item ID and cluster ID are passed through a hashing function to determine the dedicated cluster segment in the item pool. Each cluster segment consists of multiple slots to store item features, with each segment having an equal number of slots. When multiple items hash to the same slot, the later item overwrites the earlier one by design; this keeps the pool continuously refreshed with recent items and bounds its memory footprint.

\begin{algorithm}[h]
   \caption{Update engine}
   \label{alg:update}
\begin{algorithmic}
   \STATE {\bfseries Input:} Item IDs in $i$th training batch $x_i=[x_{i1}, x_{i2}, \ldots, x_{iB}]$, size $B$, slots per cluster segment $S$
   \FOR{$j=1$ {\bfseries to} $B$}
   \STATE 1. Get cluster id of $x_{ij}$: $c_{ij}$
   \STATE 2. Hash $x_{ij}$ to index $c_{ij} \cdot S + (x_{ij} \bmod S)$ in the item pool
   \ENDFOR
\end{algorithmic}
\end{algorithm}

\begin{figure*}[t]
  \centering
  \includegraphics[width=0.58\textwidth]{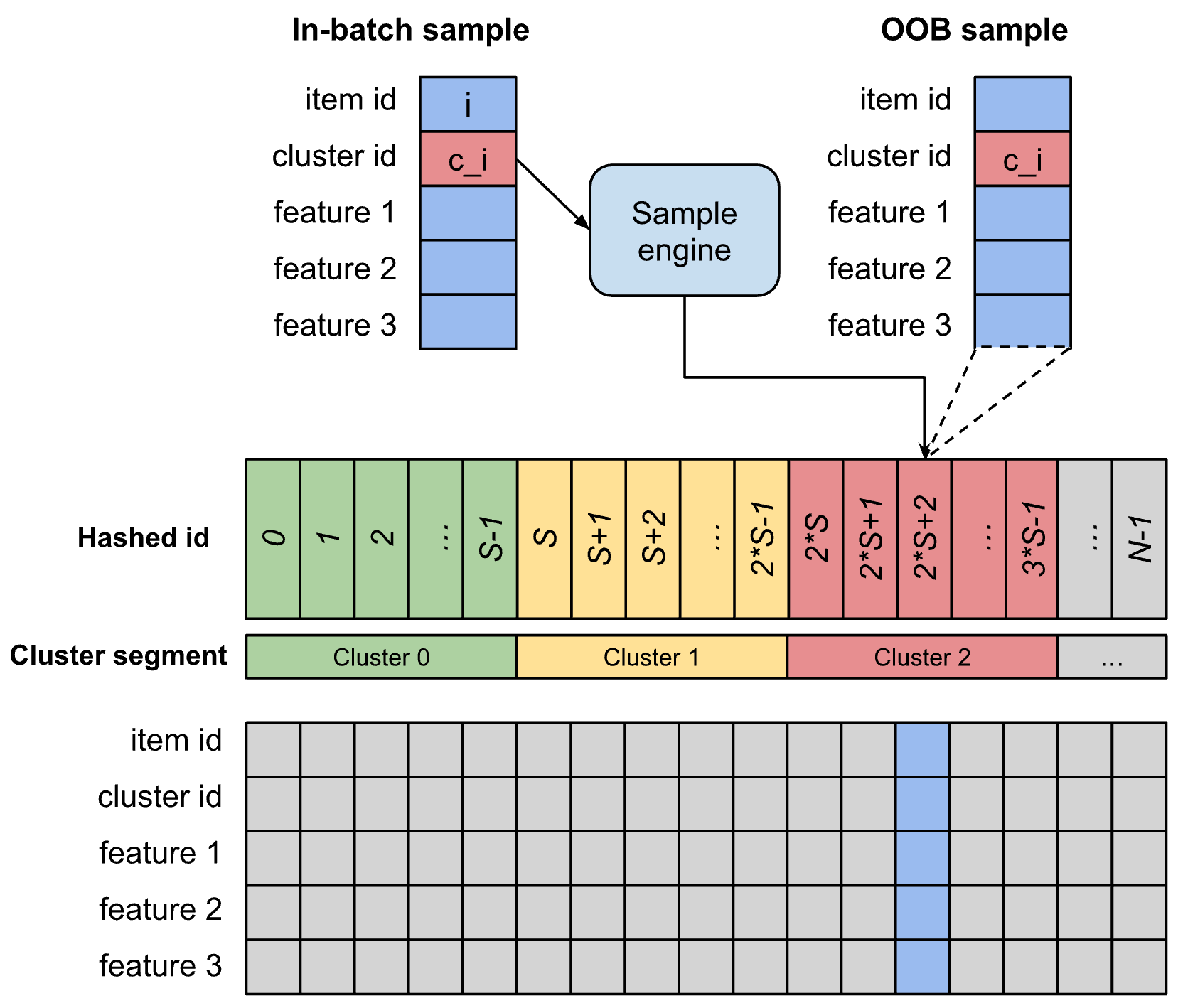}
  \caption{Cluster-based sampling}
  \label{fig:pool_sample}
\end{figure*}

During sampling, as illustrated in Figure \ref{fig:pool_sample} and Algorithm \ref{alg:sample}, the in-batch samples' cluster IDs are used to identify the target cluster segment to sample from. Within the targeted cluster segment, an existing item and its features are randomly selected to be used as OOB samples.

\begin{algorithm}[h]
   \caption{Sample engine}
   \label{alg:sample}
\begin{algorithmic}
   \STATE {\bfseries Input:} Item IDs in $i$th training batch $x_i=[x_{i1}, x_{i2}, \ldots, x_{iB}]$, size $B$, slots per cluster segment $S$
   \FOR{$j=1$ {\bfseries to} $B$}
   \STATE 1. Get cluster ids of the in-batch samples: $c_{ij}$
   \STATE 2. Random sample $r_{ij} \in$ rand$(0, S)$
   \STATE 3. Sample item $N_{ij}$ in index $c_{ij} \cdot S + r_{ij}$ from the item pool
   \ENDFOR
\end{algorithmic}
\end{algorithm}

\section{Experiments}

\begin{table*}[htbp]
    \centering
    \caption{Comparison of negative sampling methods on public
    datasets. HR@50 and HR@100 are reported. Bold indicates
    best performance. Relative improvements over Baseline are
    shown in parentheses.}
    \label{tab:public-data}
    \setlength{\tabcolsep}{5pt}
    \resizebox{\textwidth}{!}{%
    \begin{tabular}{clccccccc}
        \toprule
        \textbf{Dataset} & \textbf{Metric} &
        \textbf{Baseline} &
        \textbf{DNS} & \textbf{CBNS} &
        \textbf{ANCE} & \textbf{GOOBS} & \textbf{Cluster GOOBS} \\
        & & \textbf{(In-batch)} &
        &
        &
        & & \\
        \midrule
        \multirow{2}{*}{Movielens-1M}
          & hr@50   & .2253
            & .2298 (+2.0\%) & .2331 (+3.5\%)
            & .2380 (+5.6\%) & .2346 (+4.2\%) & \textbf{.2415 (+7.2\%)} \\
          & hr@100  & .3588
            & .3618 (+0.8\%) & .3608 (+0.6\%)
            & .3661 (+2.0\%) & .3611 (+0.7\%) & \textbf{.3682 (+2.7\%)} \\
        \midrule
        \multirow{2}{*}{Amazon-Grocery}
          & hr@50   & .0254
            & .0261 (+2.8\%) & .0271 (+6.7\%)
            & .0288 (+13.4\%) & .0279 (+10.1\%) & \textbf{.0301 (+18.5\%)} \\
          & hr@100  & .0406
            & .0419 (+3.2\%) & .0432 (+6.4\%)
            & .0451 (+11.1\%) & .0440 (+8.3\%) & \textbf{.0470 (+15.7\%)} \\
        \midrule
        \multirow{2}{*}{Amazon-Electronics}
          & hr@50   & .0084
            & .0090 (+7.1\%) & .0099 (+17.9\%)
            & .0108 (+28.6\%) & .0110 (+30.9\%) & \textbf{.0131 (+55.6\%)} \\
          & hr@100  & .0154
            & .0163 (+5.8\%) & .0176 (+14.3\%)
            & .0186 (+20.8\%) & .0190 (+23.5\%) & \textbf{.0201 (+30.2\%)} \\
        \midrule
        \multirow{2}{*}{Amazon-Home}
          & hr@50   & .0050
            & .0054 (+8.0\%) & .0061 (+22.0\%)
            & .0065 (+30.0\%) & .0067 (+34.2\%) & \textbf{.0074 (+47.3\%)} \\
          & hr@100  & .0082
            & .0088 (+7.3\%) & .0094 (+14.6\%)
            & .0099 (+20.7\%) & .0102 (+23.9\%) & \textbf{.0118 (+42.3\%)} \\
        \bottomrule
    \end{tabular}%
    }
\end{table*}

\subsection{Public Datasets}\label{public_data}
\subsubsection{Datasets}
The performance of the cluster based negative sampling is first evaluated on the public datasets MovieLens-1M and Amazon Reviews (subsets Grocery, Electronics, Home) with full-shuffle similar to previous work on recommendations \cite{jiaqi2023model, jiaqi2024hstu}. For pre-processing, timestamp-based splitting is used, ordering data by timestamps and taking the first 80\% for training and 20\% for evaluation. The rating label is converted to a binary label where rating 1--2 are negatives and 3--5 are positives. Features include user id, item id, and cluster id (e.g.\ genre id, category id) directly available from the datasets. Each user's historical interaction sequence is extracted as user features. % REVISED: R2 and R3 both noted that the public experiments use category labels
% rather than LLM-derived clusters, leaving the LLM framing unvalidated in a
% controlled setting. Reframed as a deliberate design choice, with the reason
% stated and the claim explicitly bounded.
On the public datasets we deliberately use the readily available category and genre labels as the cluster assignment, and reserve the LLM-based clustering of Section~\ref{llm} for the production deployment (Section~\ref{sec:industry}). The public corpora are small---MovieLens-1M and the Amazon Reviews subsets contain on the order of $10^4$--$10^5$ items, three to four orders of magnitude fewer than the production corpus, with only a short title and a few sparse attributes per item---so an LLM-derived partition of that space largely recovers the category taxonomy the datasets already ship with, at additional cost and with no additional information. The public experiments are therefore intended to isolate the contribution of the cluster-based sampling mechanism itself, holding the clustering source fixed at a simple and fully reproducible signal; the LLM-based clustering is evaluated as part of the complete system in the production deployment, where no curated taxonomy is available. We make no claim that LLM-based clustering outperforms category labels on the public benchmarks.

It is important to note that the evaluation protocol intentionally avoids $k$-core filtering (e.g., removing users or items with fewer than 5 or 10 interactions) and evaluates the target item against the entire global corpus rather than a small sample of 100 random negatives. As demonstrated by Krichene and Rendle~\cite{krichene2020sampled}, sampled evaluation metrics are often inconsistent with exact global ranking and artificially inflate absolute Hit Rate numbers. By retaining the full sparsity of the original datasets and performing exact global ranking, this experimental setup more accurately reflects the difficulty of real-world cold-start deployment scenarios. Consequently, while the absolute HR@50 and HR@100 values reported here are lower than those in studies employing heavy $k$-core filtering and sampled metrics, the relative performance improvements between sampling strategies remain robust and unbiased.

\subsubsection{Models and Negative Sampling Strategies}
A standard two-tower model is used as the backbone with
six negative sampling strategies implemented on top of it
to enable a comprehensive comparison with the state of the art:

\begin{itemize}
    \item \textbf{Baseline} (in-batch negative sampling
    w/\ LogQ correction): the standard industry baseline
    using in-batch negatives with LogQ correction to reduce
    over-penalization of frequent items. A false-negative
    mask is applied to suppress logits of known positives
    appearing as in-batch negatives.

    \item \textbf{DNS}~\cite{zhang2013optimizing}: Dynamic
    Negative Sampling selects negatives with the highest
    predicted relevance scores from the current model
    checkpoint, i.e., items the model currently believes
    are most relevant but are actually negative. This is the
    canonical hard negative baseline in the recommendation
    literature.

    \item \textbf{CBNS}~\cite{wang2021cross}: Cross-Batch
    Negative Sampling caches item embeddings from recent
    mini-batches and reuses them as negatives for the current
    batch. Like GOOBS, CBNS leverages out-of-batch
    items; however, it selects negatives based on recency of
    appearance rather than semantic cluster membership.

    \item \textbf{ANCE}~\cite{xiong2020approximate}:
    Approximate Nearest Neighbor Negative Contrastive
    Estimation builds a global ANN index over the entire
    item corpus and asynchronously refreshes it during
    training. Negatives are selected as the approximate
    nearest neighbors of each query in the embedding space,
    making them the most geometrically hard negatives
    available globally.

    \item \textbf{GOOBS} (Global Out-of-Batch Sampling): extends the Baseline
    with random out-of-batch (OOB) samples drawn uniformly
    from a maintained item pool in real-time. A pre-training
    pool loading step ensures high OOB hit rates from the
    start of training.

    \item \textbf{Cluster GOOBS}: augments the random OOB
    samples with hard negatives drawn from the same
    semantic cluster as the positive item, using a ratio of
    cluster to random OOB samples of 1:15 for Movielens-1M
    and 1:31 for Amazon Reviews datasets.
\end{itemize}

\subsubsection{Results}
Table~\ref{tab:public-data} reports HR@50 and HR@100 across all four datasets.
Cluster GOOBS achieves the best performance on every dataset and metric,
with gains over the in-batch baseline ranging from $+7.2\%$ (Movielens-1M HR@50)
to $+55.6\%$ (Amazon-Electronics HR@50), confirming that cluster-based hard
negatives provide a consistent and substantial training signal across diverse
recommendation domains.

Several trends are worth highlighting.
First, the benefit of out-of-batch sampling is already evident with GOOBS alone:
random OOB samples yield $+4.2\%$--$+34.2\%$ HR@50 gains over the in-batch
baseline, demonstrating that exposure to a broader item pool during training
improves generalization even without hard-negative selection.
Second, replacing random OOB samples with cluster-based hard negatives
(Cluster GOOBS) consistently outperforms all baselines, including the
geometrically hard ANCE method, which requires a global ANN index refresh.
This is particularly pronounced on the sparser Amazon datasets: on
Amazon-Electronics, Cluster GOOBS achieves $+55.6\%$ HR@50 vs.\ $+28.6\%$ for
ANCE, and on Amazon-Home, $+47.3\%$ HR@50 vs.\ $+30.0\%$ for ANCE.
Third, the gains are more pronounced on HR@50 than HR@100 across all datasets,
suggesting that cluster-based negatives particularly sharpen the model's
ability to rank the most relevant items at the top of the retrieval list ---
the regime most critical for downstream ranking stages.

Overall, these results validate that semantic cluster membership provides a
more effective hard-negative signal than recency (CBNS), dynamic score-based
selection (DNS), or approximate nearest-neighbor geometry (ANCE), while
requiring no global index and remaining compatible with real-time serving.

\subsection{Industry Datasets}\label{sec:industry}
To evaluate the effectiveness of Cluster GOOBS in a real-world application, it is deployed to the entry content-discovery surfaces of a large-scale social media application, serving live production traffic, and evaluated via online A/B testing.

Unlike public datasets, real-world datasets present greater complexity and challenges, often exhibiting popularity bias. The application serves a user base on the order of one hundred million monthly active users (MAU), with an eligible item corpus on the order of tens of millions of posts and models trained on billions of user--item interactions. Despite this scale, the top 100 items represent roughly 50\% of the total impressions. This indicates that the existing recommendation system does not have sufficient exploration on long tail items, causing users to interact with a limited set of items and creating a strong system feedback loop.

\subsubsection{Setup}
For both control and test group, 3\% of users are randomly selected respectively. In the control group, the users are served by a two-tower model architecture with GOOBS. In the test group, the users are served by the same two-tower model architecture with Cluster GOOBS. The experiment ran for 14 consecutive days on live production traffic. Full production details---including the complete model architecture, feature set, and training hyperparameters---are omitted due to proprietary constraints.

\subsubsection{Cluster Selection}
\begin{figure*}[t]
  \centering
  \includegraphics[width=0.85\textwidth]{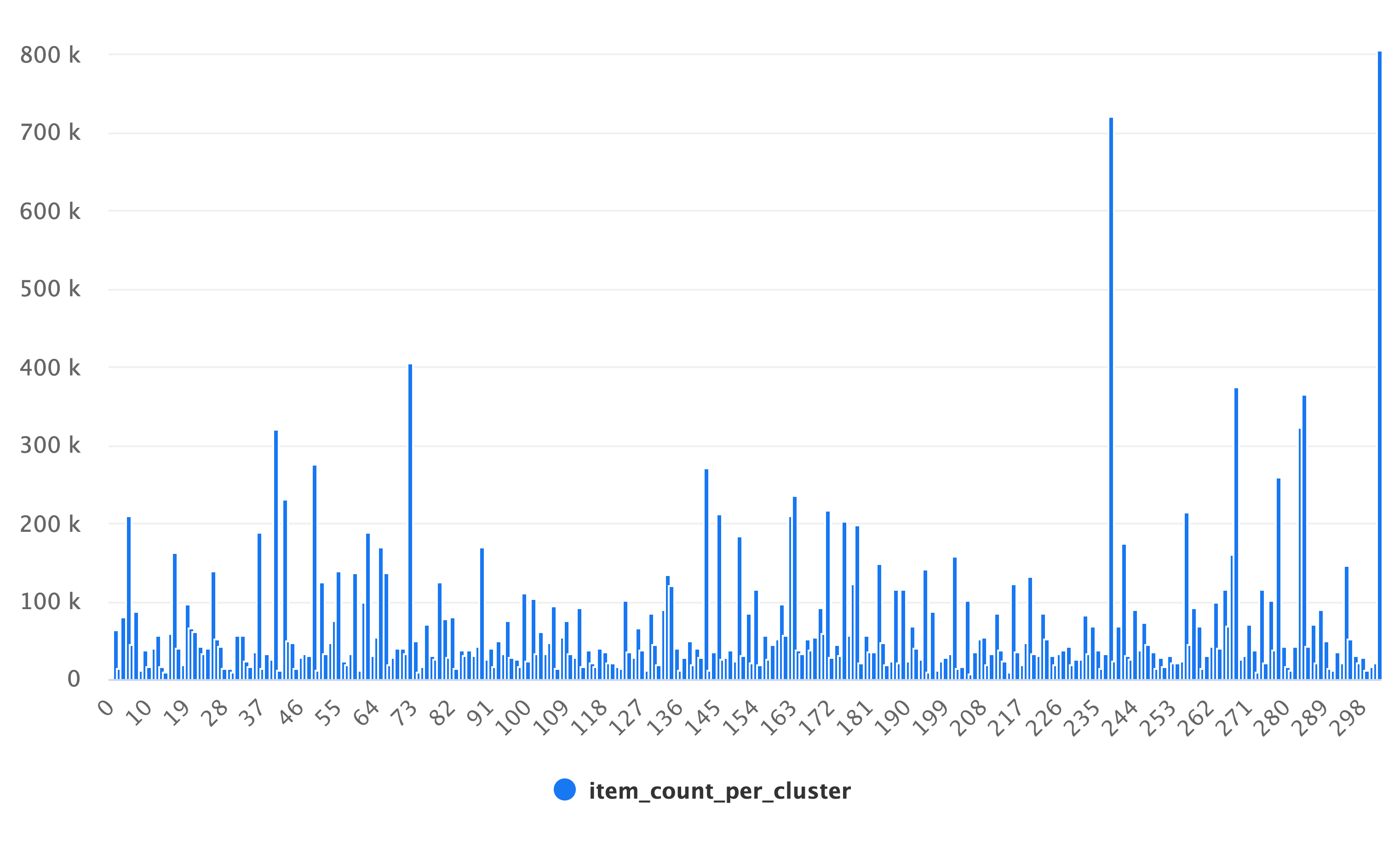}
  \caption{Cluster size distribution}
  \label{fig:cluster_size}
\end{figure*}

As discussed in Section \ref{llm}, in the real-data application, an LLM-based content-understanding model is leveraged to learn media representations and derive clusters. We compared several in-house LLM-based content features and selected the one that yielded the best cluster distribution balance and semantic clarity; this component is modular and can be replaced by any open-source LLM-based content embedding. In an offline sweep of the cluster count from 100 to 10{,}000, we found that 250--500 clusters gave the best balance between negative hardness and false-negative risk; accordingly, the deployment uses in total 300 clusters, 98\% of which have $\geq 10k$ items. Figure \ref{fig:cluster_size} summarizes the cluster size across different clusters.

\subsubsection{Results}
The models are optimizing user engagements such as clicks. Therefore, CTR (click-through-rate) is used to measure online performance. Our system comprises multiple retrieval sources. The model improvement targets one of the largest sources in production. As shown in Table \ref{table-ctr}, Cluster GOOBS improves the CTR of that source by 53\% ($p < 0.01$). Because this gain lands on the highest-traffic source in the system, it also translates into a $+6.5\%$ improvement in overall system-level CTR. Meanwhile, it only introduces minor training QPS (Queries Per Second) regression of $-1.4\%$ (with no regression for inference QPS), demonstrating the strong scalability and efficiency of the algorithm and making it easily adoptable in production.

The contents distribution is further analyzed to measure the popularity debias from Cluster GOOBS. Items having $\geq 1$K impressions in the past 1 day are used as targeted item cohorts. In the test group, the number of items in this $\geq$1K-impression cohort increased by around 50\% as shown in Table \ref{table-popular}. The impression contribution from the top 100 items dropped from 50\% to 32\%, indicating a significant improvement in popularity bias.

\begin{table}[h]
  \centering
  \caption{Online CTR comparison between different sampling techniques with the same two-tower architecture, measured over the 14-day A/B test. The numbers here are relative improvements.}
  \label{table-ctr}
  % REVISED: added the Overall CTR column so the system-level effect is
  % reported alongside the source-level one. Headers are stacked over two lines
  % (matching Table 3 below) because a single-line four-column header overflows
  % the column by ~40pt.
  \setlength{\tabcolsep}{4pt}
  \begin{tabular}{@{}lccc@{}}
    \toprule
    Model & \shortstack{Source\\CTR} & \shortstack{Overall\\CTR} & \shortstack{Training\\QPS} \\
    \midrule
    GOOBS (control) & 0\% & 0\% & 0\% \\
    Cluster GOOBS (test) & +53\% & +6.5\% & $-1.4\%$ \\
    \bottomrule
  \end{tabular}
\end{table}

\begin{table}[htbp]
  \centering
  \caption{Popularity debias comparison between different sampling techniques with the same two-tower architecture. The numbers here are relative improvements.}
  \label{table-popular}
  \setlength{\tabcolsep}{4pt}
  \begin{tabular}{@{}lcc@{}}
    \toprule
    Model & \shortstack{Imp.\ buckets\\$\geq$1k} & \shortstack{Top 100 items'\\imp.\ contrib.} \\
    \midrule
    GOOBS (control) & 0\% & 50\% \\
    Cluster GOOBS (test) & +50\% & 32\% \\
    \bottomrule
  \end{tabular}
\end{table}

\section{Conclusions}

In this paper, we presented a cluster-based hard negative sampling technique together with a practical, real-time framework that realizes it in production for two-tower retrieval. The technique draws hard negatives from semantic clusters derived from LLM-based content representations, addressing the limitations of traditional negative sampling methods, which often produce easy negatives that fail to sufficiently challenge the model.

The system utilizes an item pool to efficiently manage and update OOB samples, ensuring minimal computational complexity while handling billions of training data points. The integration of a customized hashing function within the update engine and a targeted sampling engine allows for precise and effective negative sampling, maintaining the freshness and relevance of the item pool throughout the training process.

Overall, the proposed system provides a robust and scalable solution for improving the performance of recommendation models, offering a seamless integration path into production environments. Future work may explore further optimizations and extensions, including weighted cluster-based negative sampling and user query based negative sampling.

% REVISED: acknowledgments enabled for the camera-ready.
\begin{acks}
We'd like to thank Pavitra Rengarajan, Guanfeng Liang, Claire Zhang, Henry A Lin, Kien Pham, Chenglin Lu, Kevin Zhao, Shoya Yoshida, Chris Smith, Srinath Sridhar, Tomer Bar, Mustafa Lokhandwala, Brion Spensieri, Hsiao-Ping Tseng, Qin Huang, Ziyi Zhao, Xinchen Hu, Bokai Cao, Miao Yu, Evie Feng, Ziyang Xie, Jinqiang Liu, Xianjie Chen, Wei Zheng, Aashu Singh, Michael Ge, Yuan Zeng, Sung Ha Hwang, Jianyu Wang, Shilin Ding, Xinyao Hu, Hong Yan for overall support on this project.
\end{acks}

\bibliographystyle{ACM-Reference-Format}
\bibliography{paper}

\end{document}